\documentclass[aps,pra,onecolumn,showkeys,nofootinbib,nobibnotes,superscriptaddress,10pt]{revtex4-2}%
\usepackage{amssymb}
\usepackage{amsmath}
\usepackage{amsfonts}
\usepackage{amsmath}
\usepackage{graphicx}
\usepackage{dcolumn}
\usepackage{epsfig}
\usepackage{color}
\usepackage{hyperref}
\usepackage{hyphenat} 
\usepackage{bm}
\usepackage{float}
\usepackage{amsthm}
\usepackage{enumitem}
\usepackage[portuguese]{babel}
\usepackage[latin1]{inputenc}
\usepackage[T1]{fontenc}
\usepackage{url}
\usepackage{setspace}
\usepackage{footmisc}
\usepackage[paperwidth=210mm,paperheight=297mm,centering,hmargin=2.8cm,vmargin=1.8cm]{geometry} 
\newcommand{\be}{\begin{equation}}
\newcommand{\beast}{\begin{equation*}}
\newcommand{\ee}{\end{equation}}
\newcommand{\eeast}{\end{equation*}}
\newcommand{\br}{\begin{eqnarray}}
\newcommand{\brast}{\begin{eqnarray*}}
\newcommand{\er}{\end{eqnarray}}
\newcommand{\erast}{\end{eqnarray*}}
\newcommand{\bse}{\begin{subequations}}
\newcommand{\ese}{\end{subequations}}

\newcommand{\bd}{\begin{displaymath}}
\newcommand{\ed}{\end{displaymath}}

\newcommand{\bfig}{\begin{figure}}
\newcommand{\efig}{\end{figure}}

\begin{document}
\title{Disentanglement of a bipartite system portrayed in a (3+1)D 
compact Minkowski manifold; quadridistances and quadrispeeds}
\author{Salomon S. Mizrahi}
\affiliation{Universidade Federal de São Carlos, CCET, Departamento de Física \\
Via Washington Luis 235, São Carlos, SP, Brazil}
\email{salomonsmizrahi@gmail.com}
\thanks{CNPq, contrato No. 301977/2024-0 \\ This is a revised version of chapter 8  
in the \emph{Proceedings of the Second International Workshop on Quantum 
Non-stationary Systems}, held in University of Brasília, August 28 to September 1, 2023}

\begin{abstract}
In special relativity, trajectories of particles, whether massive or massless, in 4D, can be  
displayed in the 3+1 Minkowski space-time manifold. On the other hand, in quantum mechanics,  
trajectories in phase space are not strictly defined because coordinate and linear momentum  
cannot be measured simultaneously with arbitrary precision, as these variables do not commute 
with each other. They are not sharply defined within Hilbert space formalism. 
Nonetheless, out from the density matrix representing a quantum system  
the extracted information still yields an enhanced description of its properties, and, furthermore, 
by arranging adequately the matrix one can acquire additional information from its content.  
Following these lines of conduct this paper focuses on a closely related issue, the definition and 
meaning of velocity and speed of a typical quantum phenomenon, the disentanglement for a bipartite 
system when its evolution is displayed in a 4D pseudo space-time, whose coordinates are combinations 
of the density matrix entries. The formalism is based on the definition of a compact Minkowski manifold, 
where trajectories are defined along the same reasoning of special relativity in the Minkowski manifold. 
The space-like  and time-like regions acquire different meaning, being termed entangled-like and 
separable-like, respectively. The definition and meaning of velocity and speed of disentanglement 
follow naturally from the formalism. Depending on the dynamics of the physical system state, trajectories 
may go forth and back from entanglement to separability regions of the compact Minkowski manifold. When 
the physical time $t$ is introduced as an intrinsic variable into the formalism, a phenomenon commonly 
known as sudden death occurs during irreversible evolution when a state that is initially entangled 
becomes separable.
\end{abstract}
\date{\today}
\maketitle
\tableofcontents
%
\section{Introduction}
%
According to current physical theories, it is widely admitted that the principle of causality 
cannot be violated, i.e., it is impossible for energy, matter or ``meaningful information'' to travel 
at a pace faster than the speed of light, also referred as superluminal or supercausal transmission. 
Conforming to special relativity (SR) only particles with zero mass, such as photons and possibly 
neutrinos, can travel at the speed of light $c$. Although superluminal motion of any kind 
of particle contradicts the theory of relativity, nonetheless, in quantum mechanics (QM) some 
effects suggest otherwise when submitted to the sieves of current wisdom. These include the 
tunneling effect, the non-local spooky action at a distance, and the loss of entanglement of a 
bipartite/multipartite system under measurement or environment influence \cite{EPR,salomon1}.

Historically, the connection between SR and quantum mechanics (QM) was 
established shortly after its inception in 1926. Schrödinger explored this relationship without 
publishing his findings \cite{schr}. Subsequently, the formalism was developed by Klein 
\cite{klein}, Gordon \cite{gordon}, and Dirac \cite{dirac}, leading to the Klein-Gordon 
equation for integer spin particles and the Dirac equation for spin 1/2 particles \cite{davydov}. 
Following this, the concept of quantized fields in quantum theory emerged, along with the principle 
of Lagrangian invariance under Lorentz transformations. 

I want to emphasize here that the purpose of this manuscript is not to combine non-relativistic 
QM with SR as a variant of what was already achieved up to now, the aim here consists in using, 
as motivation, the formal structure of SR to define a \emph{speed} for the disentanglement of a 
bipartite system in a Hilbert space, $\mathcal{H}^{\otimes 2}$, depending on the parameters 
that characterize it and even when there is a time dependence.  

The discussion surrounding the separability and entanglement of bipartite states was already explored 
from a novel perspective in \cite{HSS1,HSS2}. The approach enables the depiction of 
trajectories in a flat Minkowski manifold , featuring a bounded domain for the (3+1)D 
with two new coordinate systems. This differs from the space-time variables in SR, 
which are defined by the domain $\left(-\infty ,\infty \right)$. Those studies focused on two-qubit 
states, characterized by parameters present in the system's Hamiltonian and represented by a mixed 
state density operator. The theory utilizes formal similarities with special relativity, particularly 
through a restricted compact Minkowski manifold (CMM) because, as quantum mechanics gathers 
information probabilistically, as such, the measure space \((E,X,\mu)\)\footnote{Where $E$ 
is a set, $X$ is a $\sigma$ algebra of subsets of $E$ and $\mu$ is a non-negative measure on $E$, 
defined on the $X$ sets.} has measure $\mu(E) =1$. In the manifold space the trajectories meander 
through two distinct regions: one for world lines, referred to ``separable-like'', with allusion 
to the time-like region of SR, while the other is termed ``entangled-like'', corresponding to 
the space-like region of SR.

In paper \cite{HSS1} several physical systems proposed in references \cite{peres1, horo1, 
blank, werner, gisin, yu-eber1, yu-eber3, bellomo, mazzola, almeida, lopez, das} were analyzed. 
Depending on the set of parameters of each 2-qubit system, the trajectories related to each a state 
display a common behavior, starting as an entangled state, by changing the values of the parameters, 
they progressively evolve towards separable states. The respective trajectories 
can even cross the ``light-like'' line separating the two regions, oscillating back and forth.

In SR trajectories represent physical particles evolving in space-time. In quantum mechanics 
(Hilbert space), the trajectories in the CMM have a different meaning. If physical time is 
introduced into the formalism (parameters depending on time), the dynamical evolution may imply 
action at a distance without transference of meaningful information. Moreover, with a 
convenient change of coordinates it is verified that the ``speed of disentanglement'' 
exceeds the ``speed of light'', as defined in the SR context.

Recently, in line with the present theme an experiment was reported in \cite{jian}, where 
the time taken for disentanglement of states was calculated. 
It was reported that an electron ejected by a laser pulse remains entangled with the 
state of the remaining electrons of the atom. According to an author of the paper, J. Burgdörfer, 
in \cite{burg}: ``...this means the ejected electron exists in a superposition of having 
left the atom multiple times without a definitive moment. The state of the remaining 
electron affects the likelihood of the ejected electrons' birth time: if the remaining 
electron has higher energy, the ejected electron likely left earlier; if it has lower 
energy, it likely left later, averaging around 232 attoseconds.''

The paper contains a concise review of bipartite quantum systems, focusing specifically on a 
generic two-qubit mixed state. It draws a formal comparison between the SR Minkowski manifold 
and the CMM. In sequence, after defining new coordinate systems, the so-called Peres-Horodecki 
criterion is adopted to establish the distinction between separable-like and entangled-like 
regions in the manifold space. Subsequently, it becomes essential to define the velocity and 
speed of disentanglement as the system state evolves through the CMM, which is equipped with 
frames where pseudo time and the $\mathcal{R}^3$ Euclidean metric space create the \((1+3)D\) 
coordinate systems. When the common physical time $t$ is introduced 
as an internal free parameter, one can follow the trajectory evolution, even undulating, throughout 
regions of entanglement and separability. A phenomenon currently referred in the literature 
\cite{yu-eber1,yu-eber3} as \emph{sudden death} is displayed in the current formalism 
\cite{HSS1,HSS2}, where \emph{revivals} also occur.
 %
\section{The two-qubit state}
%
The most general two-qubit state (pure or mixed after tracing out over a $(N-2)D$ 
subsystem) can be expressed in the so-called Fano's form, (see, for instance. Eqs. 
(34) and (54) in \cite{GS})
\begin{equation} \label{a15}
\hat{\rho}=\frac{1}{2^{2}}\left( 1_{1}\otimes 1_{2}+1_{1}\otimes \vec{\sigma}%
_{2}\cdot \vec{P}_{2}+\vec{\sigma}_{1}\cdot \vec{P}_{1}\otimes 1_{2}+\vec{%
\sigma}_{1}\cdot \overleftrightarrow{M}\cdot \vec{\sigma}_{2}\right) 
\end{equation}
where $\vec{\sigma}$ \ is a vector whose$,$ $x,y$, and $z$ components are
the Pauli matrices $\sigma _{x}$, $\sigma _{y}$, and $\sigma _{z}$. $\vec{P}%
_{k}=\mathrm{Tr}\left( \vec{\sigma}_{k}\hat{\rho}\right) $ is a polarization
vector (PV) associated with each qubit, $k=1,2$, $\overleftrightarrow{M}$ is a 
dyadic operator encasing the correlation matrix (CM)
\begin{equation}  \label{a17}
\mathbb{M}=\left( 
\begin{array}{ccc}
M_{xx} & M_{xy} & M_{xz} \\ 
M_{yx} & M_{yy} & M_{yz} \\ 
M_{zx} & M_{zy} & M_{zz}%
\end{array}\right) ,
\end{equation}
(the subscript on the left stands for qubit 1 and the other for qubit 2) with
entries $M_{ij}=\left\langle \sigma _{1,i}\sigma _{2,j}\right\rangle =%
\mathrm{Tr}\left( \hat{\rho}\sigma _{1,i}\sigma _{2,j}\right) $ and $%
\left\vert M_{ij}\right\vert \leq 1$. A separable two-qubit state has the
following properties: (a) the PV's are written as $\vec{P}_{\mu } =  \sum_{k}p_{k}\vec{Q}_{\mu }^{\left(
k\right) }$, where $\vec{Q}_{\mu }^{\left( k\right) }$ ($\mu =1,2$) is a
vector, $\left\vert \vec{Q}_{\mu }^{\left( k\right) }\right\vert \leq 1$,
the superscript $k$ characterizes a direction in 3D and, (b) whenever the dyadic can be written as $%
\overleftrightarrow{M}=\sum_{k}p_{k}\overleftarrow{Q}_{1}^{\left( k\right) }\overrightarrow{Q}%
_{2}^{\left( k\right) }$, with weights $p_{k}\in \left[ 0,1\right] $ and $%
\sum_{k}p_{k}=1$, the state (\ref{a15}) becomes 
\begin{equation} \label{a19}
\hat{\rho}^{\mathrm{sep}}=\frac{1}{4}\sum_{k}p_{k}\left( 1_{1}+\vec{Q}%
_{1}^{\left( k\right) }\cdot \vec{\sigma}_{1}\right) \otimes \left( 1_{2}+%
\vec{Q}_{2}^{\left( k\right) }\cdot \vec{\sigma}_{2}\right) ,  
\end{equation}
a form which characterize the qubits separability state.
%
\subsection{Polarization vectors and correlation matrix}
%
The state (\ref{a15}) can also be displayed, see \cite{HSS1,HSS2}, in matrix form, 
\begin{align}
\hat{\rho}& =\frac{1}{4}\left( 
\begin{array}{lr}
1+P_{1,z}+P_{2,z}+M_{zz} & P_{2,x}-iP_{2,y}+M_{zx}-iM_{zy} \\ 
P_{2,x}+iP_{2,y}+M_{zx}+iM_{zy} & 1+P_{1,z}-P_{2,z}-M_{zz} \\ 
P_{1,x}+iP_{1,y}+M_{xz}+iM_{yz} & M_{xx}+M_{yy}-i\left( M_{xy}-M_{yx}\right) 
\\ 
M_{xx}-M_{yy}+i\left( M_{xy}+M_{yx}\right)  & P_{1,x}+iP_{1,y}-M_{xz}-iM_{yz}%
\end{array}%
\right.   \notag \\
& \left. 
\begin{array}{lr}
P_{1,x}-iP_{1,y}+M_{xz}-iM_{yz} & M_{xx}-M_{yy}-i\left( M_{xy}+M_{yx}\right) 
\\ 
M_{xx}+M_{yy}+i\left( M_{xy}-M_{yx}\right)  & P_{1,x}-iP_{1,y}-M_{xz}+iM_{yz}
\\ 
1-P_{1,z}+P_{2,z}-M_{zz} & P_{2,x}-iP_{2,y}-M_{zx}+iM_{zy} \\ 
P_{2,x}+iP_{2,y}-M_{zx}-iM_{zy} & 1-P_{1,z}-P_{2,z}+M_{zz}%
\end{array}%
\right) ,  \label{a21}
\end{align}
that depends on fifteen free parameters, with constraint $\mathrm{Tr}%
 \hat{\rho} ^{2}\leq 1$. The polarization vectors are
\begin{subequations} \label{a23}
\begin{align}
\mathbb{P}_{1}& =\left( 
\begin{array}{ccc}
P_{1,x} & P_{1,y} & P_{1,z}%
\end{array}%
\right) ^{\intercal }=\left( 
\begin{array}{ccc}
2\Re\left( \rho _{13}+\rho _{24}\right)  & -2\Im\left( \rho
_{13}+\rho _{24}\right)  & 2\left( \rho _{11}+\rho _{22}\right) -1%
\end{array}%
\right) ^{\intercal }  
\label{a23a} 
\\
\mathbb{P}_{2}& =\left( 
\begin{array}{ccc}
P_{2,x} & P_{2,y} & P_{2,z}%
\end{array}%
\right) ^{\intercal }=\left( 
\begin{array}{ccc}
2\Re\left( \rho _{12}+\rho _{34}\right)  & -2\Im\left( \rho
_{12}+\rho _{34}\right)  & 2\left( \rho _{11}+\rho _{33}\right) -1%
\end{array}%
\right) ^{\intercal},  
\label{a23c}
\end{align}
\end{subequations}
the superscript $^{\intercal}$ stands for transposition, $\Re$ for the real part of 
the argument, $\Im$ for the imaginary part, and the CM (\ref{a17}) can be expressed 
as 
\begin{equation}
\mathbb{M}=\left( 
\begin{array}{ccc}
2\Re\left( \rho _{14}+\rho _{23}\right)  & 2\Im\left( \rho
_{23}+\rho _{41}\right)  & 2\Re\left( \rho _{13}-\rho _{24}\right)  \\ 
2\Im\left( \rho _{41}+\rho _{32}\right)  & 2\Re\left( \rho
_{23}-\rho _{14}\right)  & 2\Im\left( \rho _{24}-\rho _{13}\right)  \\ 
2\Re\left( \rho _{12}-\rho _{34}\right)  & 2\Im\left( \rho
_{34}-\rho _{12}\right)  & 1-2\left( \rho _{22}+\rho _{33}\right) 
\end{array}%
\right) .  \label{a25}
\end{equation}
The reduced density matrix for each qubit only depends on its own
polarization vector
\begin{subequations} \label{a27}
\begin{eqnarray}
\mathrm{Tr}_{2}\left( \hat{\rho}\right)  &=&\hat{\rho}^{_{\left( 1\right)
}}=\left( 
\begin{array}{cc}
\rho _{11}+\rho _{22} & \rho _{13}+\rho _{24} \\ 
\rho _{31}+\rho _{42} & \rho _{33}+\rho _{44}%
\end{array}%
\right) =\frac{1}{2}\left( 
\begin{array}{cc}
1+P_{1,z} & P_{1,x}-iP_{1,y} \\ 
P_{1,x}+iP_{1,y} & 1-P_{1,z}%
\end{array}%
\right)   \label{a27a}
\\
\mathrm{Tr}_{1}\left( \hat{\rho}\right)  &=&\hat{\rho}^{_{\left( 2\right)
}}=\left( 
\begin{array}{cc}
\rho _{11}+\rho _{33} & \rho _{12}+\rho _{34} \\ 
\rho _{21}+\rho _{43} & \rho _{22}+\rho _{44}%
\end{array}%
\right) =\frac{1}{2}\left( 
\begin{array}{cc}
1+P_{2,z} & P_{2,x}-iP_{2,y} \\ 
P_{2,x}+iP_{2,y} & 1-P_{2,z}%
\end{array}%
\right)   \label{a27b}
\end{eqnarray}
\end{subequations}
or, $\hat{\rho}^{\left( k\right) }=\frac{1}{2}\left( \hat{1}+\vec{\sigma}%
_{k}\cdot \vec{P}_{k}\right) $, $k=1,2$ with $\left\vert \vec{P}%
_{k}\right\vert \leq 1$, correlation of one qubit with its partner 
appears in matrix (\ref{a25}). The qubits may be devoided of polarization, 
and still being correlated.
%
\subsection{Positivity Partial Transposition (PPT)}
%
A density matrix, whether full or reduced (obtained by tracing over a subset of a physical 
system's degrees of freedom), contains all the necessary information that can be extracted 
through appropriate operations. In 1996, a clever procedure was independently proposed by 
A. Peres and the Horodecki family \cite{peres1,horo1}. This method involves a specific 
shuffling of the entries in matrix (\ref{a21}). It entails performing a particular local 
operation on one qubit, which results in a different matrix, denoted as $\hat{\rho}^{T}$. 
According to the proposed Peres-Horodecki criterion (PHC) \cite{peres1,horo1}, it is possible 
to identify an entangled state when at least one eigenvalue of $\hat{\rho}^{T}$ becomes 
negative for a specific value (or set of values) of the parameters. This feature 
indicates that when the parameters take on certain particular values, the qubits possess 
some degree of entanglement. The PHC operates based on a \textit{positive partial transposition} 
(PPT) method. This involves transposing only one qubit in the original matrix, which leads 
to a reordering of the entries. After this, the positivity condition of the eigenvalues 
is assessed.

Symbolically, 
$\hat{\rho}^{T}=\left( \hat{1}_{1}\otimes \hat{T}_{2}\right)$, where $\hat{T}_{2}$ 
stands for the transposition operation on qubit $2$, for instance. Partitioning matrix (\ref{a21}) 
into four $2\times 2$ sub-blocks, the transposition is done on the diagonal entries within each 
sub-block, $\rho _{12}\rightleftarrows \rho _{21}$, $\rho _{14}\rightleftarrows \rho _{23}$, 
$\rho _{32}\rightleftarrows \rho_{41}$, $\rho _{34}\rightleftarrows \rho _{43}$, which is a 
\textit{positive map} but not completely positive. For that reason it provides \emph{a necessary 
and sufficient condition} for testing the qubits separability. Thence, it becomes possible to
characterize the separability (or entanglement) in $\hat{\rho}$ using this procedure.

\begin{figure}[hbt]
\centering
\includegraphics[width=3.5in,height=2.2in]{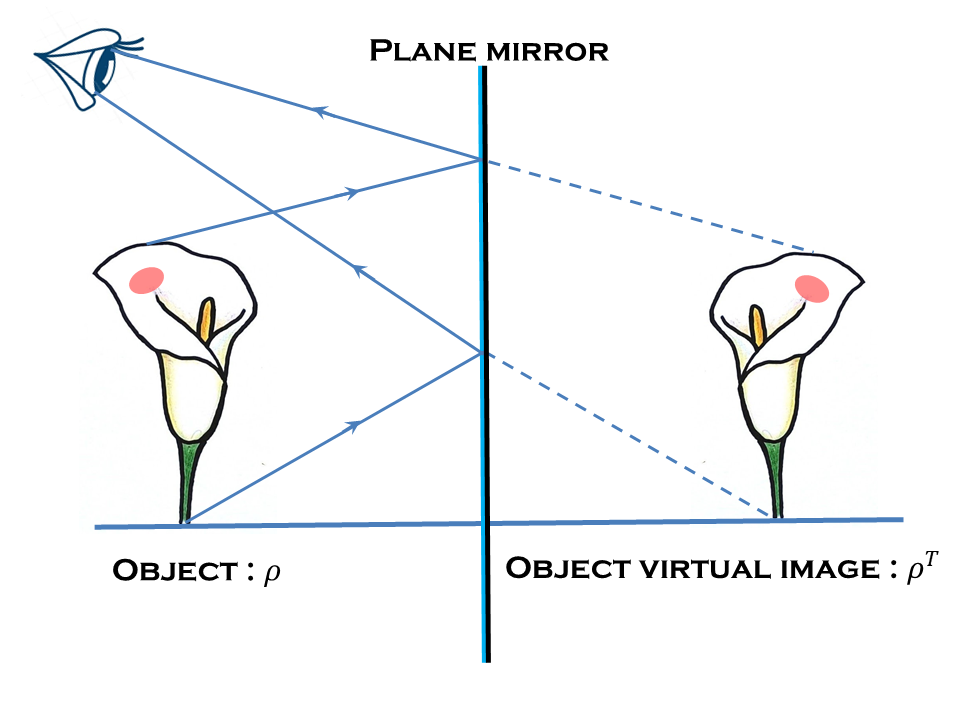} 
\caption{The stain on the inner side of the petal is only visible to an external 
observer when it is reflected in a mirror. This drawing serves as an allegorical 
illustration of the hidden information in the "object" \(\hat{\rho}\) and its 
revelation in the virtual image \(\hat{\rho}^{T}\), which is not visible to 
the observer.}
\label{fig0}
\end{figure}

The changes of the position entries in matrix (\ref{a21}), that results in 
matrix $\hat{\rho}^{T}$, offers a geometric representation in a 3D Euclidean space 
$\mathcal{R}^{3}$, resulting in a signal change of polarization vector 
$\vec{P}_{2}$, as well to a signal change in the CM (\ref{a17}),
\begin{equation}
\mathbb{P}_{2}\longrightarrow \mathbb{P}_{2}^{T}=\left( 
\begin{array}{ccc}
P_{2,x} & -P_{2,y} & P_{2,z}%
\end{array}%
\right) ^{T},  
\label{a33}
\end{equation}
and 
\begin{equation}
\mathbb{M}\longrightarrow \mathbb{M}^{T}=\left( 
\begin{array}{ccc}
M_{xx} & -M_{xy} & M_{xz} \\ 
M_{yx} & -M_{yy} & M_{yz} \\ 
M_{zx} & -M_{zy} & M_{zz}%
\end{array}%
\right)\ .  
\label{a35}
\end{equation}
Expressions (\ref{a33}) and (\ref{a35}) cannot be obtained through a unitary 
transformation; they display a reflection of the Pauli vector $\vec{\sigma}_{2}$ 
across the $x-z$ plane in $\mathcal{R}^{3}$, resulting in a virtual image\footnote
{In optics, a real image can be projected onto a surface 
because the rays converge, while a virtual image cannot be projected since the rays only 
appear to diverge.}. The drawing in Fig. \ref{fig0} features an allegorical picture  
for states $\hat{\rho}$ and $\hat{\rho}^{T}$. 
While $\hat{\rho}$ conveys standard information about physical reality, $\hat{\rho}^{T}$ 
reveals complementary information that cannot be directly extracted from $\hat{\rho}$, 
as illustrated by the red spot.
%
\subsection{\texorpdfstring{$D-7$}{TEXT} manifold class matrix with seven free parameters}
%
Greater clarity of entanglement and separability description is attained by reducing 
the number of free parameters to seven rather than keeping the fifteen in (\ref{a21}), this reduction 
make up the $D-7$ \emph{manifold class} of matrices. Assuming that the PV's are oriented along the 
$z$-direction in $\mathcal{R}^3$, 
represented as $\mathbb{P}_{k}=\left( \begin{array}{ccc} 0 & 0 & P_{k,z} \end{array} \right) ^{\intercal}$, 
for $k=1,2$, the number of non-zero entries (free parameters) in the CM is reduced to just five,
\begin{equation}
\mathbb{M}=\left( 
\begin{array}{ccc}
M_{xx} & M_{xy} & 0 \\ 
M_{yx} & M_{yy} & 0 \\ 
0 & 0 & M_{zz}%
\end{array}%
\right) ,  \label{a37}
\end{equation}
such that the matrix to be dealt with becomes
\begin{align}
\hat{\rho}_{D7} & =\frac{1}{4}\left( 
\begin{array}{cc}
1+P_{1,z}+P_{2,z}+M_{zz} & 0 \\ 
0 & 1+P_{1,z}-P_{2,z}-M_{zz} \\ 
0 & M_{xx}+M_{yy}+i\left( M_{yx}-M_{xy}\right) \\ 
M_{xx}-M_{yy}+i\left( M_{yx}+M_{xy}\right) & 0%
\end{array}%
\right.  \notag \\
& \left. 
\begin{array}{cc}
0 & M_{xx}-M_{yy}-i\left( M_{yx}+M_{xy}\right) \\ 
M_{xx}+M_{yy}-i\left( M_{yx}-M_{xy}\right) & 0 \\ 
1-P_{1,z}+P_{2,z}-M_{zz} & 0 \\ 
0 & 1-P_{1,z}-P_{2,z}+M_{zz}%
\end{array}%
\right) .  \label{a39}
\end{align}
The seven free parameters can now be combined to define a new set of eight 
parameters,
\begin{subequations} \label{a41}
\begin{align}
t_{\pm }& =\frac{1\pm M_{zz}}{2},\quad u_{\pm }=\frac{P_{1,z}\pm P_{2,z}}{2},
\label{a41a} \\
v_{\pm }& =\frac{M_{xx}\pm M_{yy}}{2},\quad w_{\pm }=\frac{M_{yx}\pm M_{xy}}{%
2}.  \label{a41b}
\end{align}%
\end{subequations}
(the digit $1$ in $t_{\pm }$ completes the set) to be treated as coordinates of frames 
contructed in the CMM. The eigenvalues of matrix (\ref{a39}) are 
\begin{equation}
\lambda _{1}=\left( t_{-}+X_{1}\right) /2,\, \lambda _{2}=\left(
t_{-}-X_{1}\right) /2,\, \lambda _{3}=\left( t_{+}+X_{2}\right) /2,\,
\lambda _{4}=\left( t_{+}-X_{2}\right) /2,
\end{equation}
where 
\begin{equation}
X_{1}^{2}=u_{-}^{2}+v_{+}^{2}+w_{-}^{2}\text{\qquad and\qquad }%
X_{2}^{2}=u_{+}^{2}+v_{-}^{2}+w_{+}^{2},  \label{a43}
\end{equation}
are specific quadratic distances from the origin in the frame system in $\mathcal{R}^3$. 
According to the PHC, the partial transposition on qubit 2 turns out to be equivalent 
in making the changes 
\begin{equation} (t_{\pm },u_{\pm },v_{\pm },w_{\pm })\rightarrow
\left( t_{\pm },u_{\pm },v_{\mp },w_{\mp }\right) 
\end{equation}
and the eigenvalues of the partially transposed matrix $\hat{\rho}^{T}_{D7}$ are 
\begin{equation}
\lambda _{1}^{T}=
\left( t_{-}+X_{1}^{T}\right) /2,\, \lambda _{2}^{T}=\left( t_{-}-X_{1}^{T}\right) /2,\, 
\lambda _{3}^{T}=\left( t_{+}+X_{2}^{T}\right) /2,\, \lambda _{4}^{T}=\left(
t_{+}-X_{2}^{T}\right) /2,
\end{equation}
where, differently from the quadratic forms (\ref{a43}), one has
\begin{equation} \label{a45}
\left( X_{1}^{T}\right) ^{2}=u_{-}^{2}+v_{-}^{2}+w_{+}^{2}\text{,\qquad
and \qquad }\left( X_{2}^{T}\right) ^{2}=u_{+}^{2}+v_{+}^{2}+w_{-}^{2}.
\end{equation}
Comparing Eqs. (\ref{a45}) with Eqs. (\ref{a43}), one observes that only parameters 
$v$ and $w$ have their subscripts signs, $+$ and $-$, interchanged. The set of eigenvalues 
$\left\{ \lambda _{i}^{T}\right\} $
can be obtained directly from the set $\left\{ \lambda _{i}\right\} $ after
doing the changes $P_{2,y}\rightarrow -P_{2,y}$ and $ M_{ky}\rightarrow
-M_{ky}\ (k=x,y,z)$ or equivalently, $ \left\{ v_{\pm }, w_{\pm } \right\} \rightarrow  
\left\{v_{\mp }, w_{\mp } \right\} $. 
%
\subsubsection{Quadridistance, or quadrilengths, in compact Minkowski manifold (CMM)}
%
Quadratic quadridistances in CMM, associated with $\hat{\rho_{D7}}$ and $\hat{\rho}^{T}_{D7}$, 
are defined as
\begin{subequations}\label{a46}
\begin{eqnarray}
s_{1}^{2} &=& t_{-}^{2}-X_{1}^{2}  \label{a46a} \\
\left( s_{1}^{T}\right) ^{2} &=& t_{-}^{2}-\left( X_{1}^{T}\right) ^{2}, \label{a46b} \\
s_{2}^{2} &=&t_{+}^{2}-X_{2}^{2},  \label{a46c} \\
\left( s_{2}^{T}\right) ^{2} &=&t_{+}^{2}-\left( X_{1}^{T}\right) ^{2},  \label{a46d}
\end{eqnarray}
\end{subequations}
and the invariance of the sums , $s_{1}^{2}+s_{2}^{2}=\left(
s_{1}^{T}\right) ^{2}+\left( s_{2}^{T}\right) ^{2}$ is verified, a symmetry that remits 
to the invariance under Lorentz transformations in SR, although operating under dissimilar 
physical condition. While each term on the left side of the equation is non-negative, 
one of the terms on the right side can be negative, which serves as an indicator of 
states entanglement.

One is inclined to consider each parameter $t_{\pm}$ as equivalent to \emph{time} as 
it occurs with SR, while $\{u_{\pm}, v_{\pm}, w_{\pm}\}$ are to be viewed as 
coordinates in $\mathcal{R}^3$ space. Despite this comparison, it is important to remind  
that the manifold has compact support, $s_i^2 \in [0, 1]$. Using SR terminology, we can 
ascribe $s_1^2 $ and $s_2^2$ as being \emph{time-like}, and they do not provide 
information regarding separability or entanglement of the qubits states. 
However, according to the PHC applied to the PT state, Eqs. (\ref{a46b}) and (\ref{a46d}) 
enclose the necessary and sufficient information to confirm whether the qubits are entangled, 
or not. If the qubits are in an entangled state for specific numerical values of the parameters,  
either $\left( s_{1}^{T} \right)^{2}$ or $\left( s_{2}^{T} \right)^{2}$) will display   
negative numbers. By using the terminology of SR, we can say that the state of the qubits 
resides in the \emph{entangled-like} region. On the other hand, when both conditions, 
$(s_{1}^{T})^{2} > 0$ and $(s_{2}^{T})^{2} > 0$, are satisfied, the state of the 
qubits is within the \emph{separable-like} region. The equality $(s_{k}^{T})^{2} = 0$ 
defines the boundary between these regions, corresponding to the surface of the 
\emph{light-like} cone in SR.
%
\section{Velocity of disentanglement}
%
The CMM coordinates in Eqs. (\ref{a46}) are related to the entries of matrix 
(\ref{a39}) as
\begin{subequations} \label{b0}
\begin{eqnarray}
s_{1} &\Longrightarrow &\left( t_{-},u_{-},v_{+},w_{-}\right) =\left( \frac{%
1-M_{zz}}{2},\frac{P_{1,z}-P_{2,z}}{2},\frac{M_{xx}+M_{yy}}{2},\frac{%
M_{yx}-M_{xy}}{2}\right) ,  \label{b0a} \\
s_{2} &\Longrightarrow &\left( t_{+},u_{+},v_{-},w_{+}\right) =\left( \frac{%
1+M_{zz}}{2},\frac{P_{1,z}+P_{2,z}}{2},\frac{M_{xx}-M_{yy}}{2},\frac{%
M_{yx}+M_{xy}}{2}\right) ,  \label{b0b} \\
s_{^{1}}^{T} &\Longrightarrow &\left( t_{-},u_{-},v_{-},w_{+}\right) =\left( 
\frac{1-M_{zz}}{2},\frac{P_{1,z}-P_{2,z}}{2},\frac{M_{xx}-M_{yy}}{2},\frac{%
M_{yx}+M_{xy}}{2}\right) ,  \label{b0c} \\
s_{2}^{T} &\Longrightarrow &\left( t_{+},u_{+},v_{+},w_{-}\right) =\left( 
\frac{1+M_{zz}}{2},\frac{P_{1,z}+P_{2,z}}{2},\frac{M_{xx}+M_{yy}}{2},\frac{%
M_{yx}-M_{xy}}{2}\right),  \label{b0d}
\end{eqnarray}%
\end{subequations}
and each component in $\left( t_{-},u_{-},v_{+},w_{-}\right)$ and 
$\left( t_{+},u_{+},v_{-},w_{+}\right)$ may depend, intrinsically, on the 
parameters present in the Hamiltonian or in the density matrix of the physical 
system. In $\mathcal{R}^3$ two ``velocities'' can be defined, 
\begin{subequations} \label{b1}
\begin{eqnarray}
\vec{V}_{1} &=&\frac{d\vec{X}_{1}}{dt_{-}}=\left( \frac{du_{-}}{dt_{-}},%
\frac{dv_{+}}{dt_{-}},\frac{dw_{-}}{dt_{-}}\right)   \label{b1a} \\
\vec{V}_{2} &=&\frac{d\vec{X}_{2}}{dt_{+}}=\left( \frac{du_{+}}{dt_{+}},%
\frac{dv_{-}}{dt_{+}},\frac{dw_{+}}{dt_{+}}\right).   \label{b1b}
\end{eqnarray}
\end{subequations}
For a dynamical system whose evolution is measured in real time $t$, as  
intrinsic parameter, and all other parameters fixed, the velocities are 
\begin{subequations} \label{b7}
\begin{eqnarray}
\left( \frac{du_{-}}{dt_{-}},\frac{dv_{+}}{dt_{-}},\frac{dw_{-}}{dt_{-}}%
\right)  &\Longrightarrow &\left( \frac{\partial u_{-}/\partial t}{\partial
t_{-}/\partial t},\frac{\partial v_{+}/\partial t}{\partial t_{-}/\partial t}%
,\frac{\partial w_{-}/\partial t}{\partial t_{-}/\partial t}\right) _{%
\mathrm{fixed}\text{ }\mathrm{parameters}},  \label{b7a} \\
\left( \frac{du_{+}}{dt_{+}},\frac{dv_{-}}{dt_{+}},\frac{dw_{-}}{dt_{+}}%
\right)  &\Longrightarrow &\left( \frac{\partial u_{+}/\partial t}{\partial
t_{+}/\partial t},\frac{\partial v_{-}/\partial t}{\partial t_{+}/\partial t}%
,\frac{\partial w_{+}/\partial t}{\partial t_{+}/\partial t}\right) _{%
\mathrm{fixed}\text{ }\mathrm{parameters}}.  \label{b7b}
\end{eqnarray}%
\end{subequations}
and the speeds being  
\begin{subequations} \label{b3}
\begin{eqnarray}
V_{1} &=&\left\vert \vec{V}_{1}\right\vert =\left( \left( \frac{du_{-}}{%
dt_{-}}\right) ^{2}+\left( \frac{dv_{+}}{dt_{-}}\right) ^{2}+\left( \frac{%
dw_{-}}{dt_{-}}\right) ^{2}\right) ^{1/2}  \label{b3a} \\
V_{2} &=&\left\vert \vec{V}_{2}\right\vert =\left( \left( \frac{du_{+}}{%
dt_{+}}\right) ^{2}+\left( \frac{dv_{-}}{dt_{+}}\right) ^{2}+\left( \frac{%
dw_{+}}{dt_{+}}\right) ^{2}\right) ^{1/2}.  \label{b3b}
\end{eqnarray}
\end{subequations}
Each speed $V_{1}$ ($V_{2}$) (for each set of internal parameter) can be
plotted in parametric form $V_{1}\left( t\right) \times t_{-}\left(
t\right) $ and $V_{2}\left( t\right) \times t_{+}\left( t\right) $. In the CMM, 
the quadrispeeds are positive 
\begin{subequations} \label{b5}
\begin{eqnarray}
\frac{ds_{1}}{dt_{-}}=\sqrt{1-V_{1}^{2}}>0,  \label{b5a} \\
\frac{ds_{2}}{dt_{+}}=\sqrt{1-V_{2}^{2}}>0  \label{b5b}\ .
\end{eqnarray}
\end{subequations}
for $V_{1}^{2}<1$ and $V_{2}^{2}<1$. $V_{1}^{2} = 1$ and $V_{2}^{2} = 1$ 
mean a pseudo ``maximum speed'', as like as the speed of light in cosmic vacuum, 
although it there is no definition of a maximal speed in the CMM.

The same goes for the PT matrix, the velocities are
\begin{subequations} \label{b9}
\begin{eqnarray}
\vec{V}_{1}^{T} &=&\frac{d\vec{X}_{1}^{T}}{dt_{-}}=\left( \frac{du_{-}}{%
dt_{-}},\frac{dv_{-}}{dt_{-}},\frac{dw_{+}}{dt_{-}}\right) ,  \label{b9a} \\
\vec{V}_{2}^{T} &=&\frac{d\vec{X}_{2}^{T}}{dt_{+}}=\left( \frac{du_{+}}{%
dt_{+}},\frac{dv_{+}}{dt_{+}},\frac{dw_{-}}{dt_{+}}\right),   \label{b9b}
\end{eqnarray}
\end{subequations}
the speeds are
\begin{subequations} \label{b11}
\begin{eqnarray}
V_{1}^{T} &=&\left\vert \vec{V}_{1}^{T}\right\vert =\left( \left( \frac{%
du_{-}}{dt_{-}}\right) ^{2}+\left( \frac{dv_{-}}{dt_{-}}\right) ^{2}+\left( 
\frac{dw_{+}}{dt_{-}}\right) ^{2}\right) ^{1/2},  \label{b11a} \\
V_{2}^{T} &=&\left\vert \vec{V}_{2}^{T}\right\vert =\left( \left( \frac{%
du_{+}}{dt_{+}}\right) ^{2}+\left( \frac{dv_{+}}{dt_{+}}\right) ^{2}+\left( 
\frac{dw_{-}}{dt_{+}}\right) ^{2}\right) ^{1/2},  \label{b11b}
\end{eqnarray}
\end{subequations}
and the squared quadrispeeds are 
\begin{subequations} \label{b13}
\begin{eqnarray}
\left( \frac{ds_{1}^{T}}{dt_{-}}\right) ^{2} = 
1-\left( V_{1}^{T}\right) ^{2}  \label{b13a} \\
\left( \frac{ds_{2}^{T}}{dt_{+}}\right) ^{2} =
1-\left( V_{2}^{T}\right) ^{2}.  \label{b13b}
\end{eqnarray}
\end{subequations}
Factually, the sets of coordinates $\left( t_{-},u_{-},v_{-},w_{+}\right)$ and 
$\left( t_{+},u_{+},v_{+},w_{-}\right)$ depend on the system's parameters, and 
for a physical system evolving in real time the equivalent to Eqs. (\ref{b7}) 
the velocities are 
\begin{subequations} \label{b15}
\begin{eqnarray}
\left( \frac{du_{-}}{dt_{-}},\frac{dv_{-}}{dt_{-}},\frac{dw_{+}}{dt_{-}}%
\right)^T &\Longrightarrow &\left( \frac{\partial u_{-}/\partial t}{\partial
t_{-}/\partial t},\frac{\partial v_{-}/\partial t}{\partial t_{-}/\partial t}%
,\frac{\partial w_{+}/\partial t}{\partial t_{-}/\partial t}\right) _{%
\mathrm{fixed parameters}}^T\ ,  \label{b15a} \\
\left( \frac{du_{+}}{dt_{+}},\frac{dv_{+}}{dt_{+}},\frac{dw_{-}}{dt_{+}}%
\right)^T &\Longrightarrow &\left( \frac{\partial u_{+}/\partial t}{\partial
t_{+}/\partial t},\frac{\partial v_{+}/\partial t}{\partial t_{+}/\partial t}%
,\frac{\partial w_{-}/\partial t}{\partial t_{+}/\partial t}\right) _{%
\mathrm{fixed parameters}}^T\ .  \label{b15b}
\end{eqnarray}
\end{subequations}
Here too, speeds can be plotted in parametric form $V_{1}^{T}\left( t\right) 
\times t_{-}\left( t\right) $ and $V_{2}^{T}\left( t\right) \times t_{+}\left( t\right)$.
%
\section{The paradigmatic Blank-Exner-Werner state model as a test track}
%
I will now illustrate the key point of this manuscript defining the ``speed of 
disentanglement'' in the CMM, and then compare its meaning to the definition 
in SR, as presented in Appendix \ref{appA}. For sake of clarity and to restrict 
the matter to simple and straightforward calculations, I utilize the well-known 
Blank-Exner-Werner (BEW) two-qubit mixed state \cite{blank,werner} that depends 
on a single parameter $x$ and, as such, the polarization vectors are null while the 
CM elements depend linearly on it 
\begin{center}
$%
\begin{tabular}{|c||c|c|c|c|c|c|c|}
\hline
{Entries} & ${P}_{1,z}$ & ${P}_{2,z}$ & ${M}_{xx}$
& ${M}_{yy}$ & ${M}_{xy}$ & ${M}_{yx}$ & ${M}%
_{zz}$ \\ \hline\hline
\multicolumn{1}{|l||}{BEW} & ${0}$ & ${0}$ & ${-x}$ & ${-x}$ 
& ${0}$ & ${0}$ & ${-x}$ \\ \hline
\end{tabular} \ . %
$
\end{center} 
The BEW state is a mixture of a 2-qubit singlet state (as like as in a two 
spin-1/2 particles), $\left\vert \psi _{-}\right\rangle$, balanced with the 
unit operator $I$ (to be represented by a $4 \times 4$ the trivial doubly stochastic 
matrix, the unit matrix), 
\begin{equation}
{\rho}\left( x\right) =x\left\vert \psi _{-}\right\rangle \left\langle
\psi _{-}\right\vert +\frac{1-x}{4}I\ ,  \label{c0}
\end{equation}
and \( x \in [0, 1] \) represents a weight or a probability. If 
we consider \( x \) as a function of time, denoted \( x(t) \), which varies 
due to the interaction between qubits with environment, we can intuitively assign the 
initial value \( x(0) = 1 \) and the limit value goes as \( \lim_{t \to \infty} x(t) = 0 \). 
This suggests that the initial pure state will inevitably transition into a 
stochastic state and, if it becomes continuously measured, the singlet state will 
eventually cease to exist. 

In the computational basis the stochastic 2-qubit state $I$ is represented as 
\begin{equation}
I=\left\vert 00\right\rangle \left\langle 00\right\vert +\left\vert
11\right\rangle \left\langle 11\right\vert +\left\vert 01\right\rangle
\left\langle 01\right\vert +\left\vert 10\right\rangle \left\langle
10\right\vert \ .   \label{c2}
\end{equation}
The expression can be factored to show its separability, $I = I_1 \otimes I_2$ 
where $I_k = \left( \left\vert 0\right\rangle \left\langle 0\right\vert + 
\left\vert 1\right\rangle \left\langle 1\right\vert \right)_k$, and
\begin{equation}
\left\vert \psi _{- }\right\rangle =\frac{\left\vert 01\right\rangle -
\left\vert 10\right\rangle }{\sqrt{2}},  \label{c6}
\end{equation}
is the singlet state. 

\begin{figure}[hbt]
\centering
\includegraphics[width=3.8in,height=2.6in]{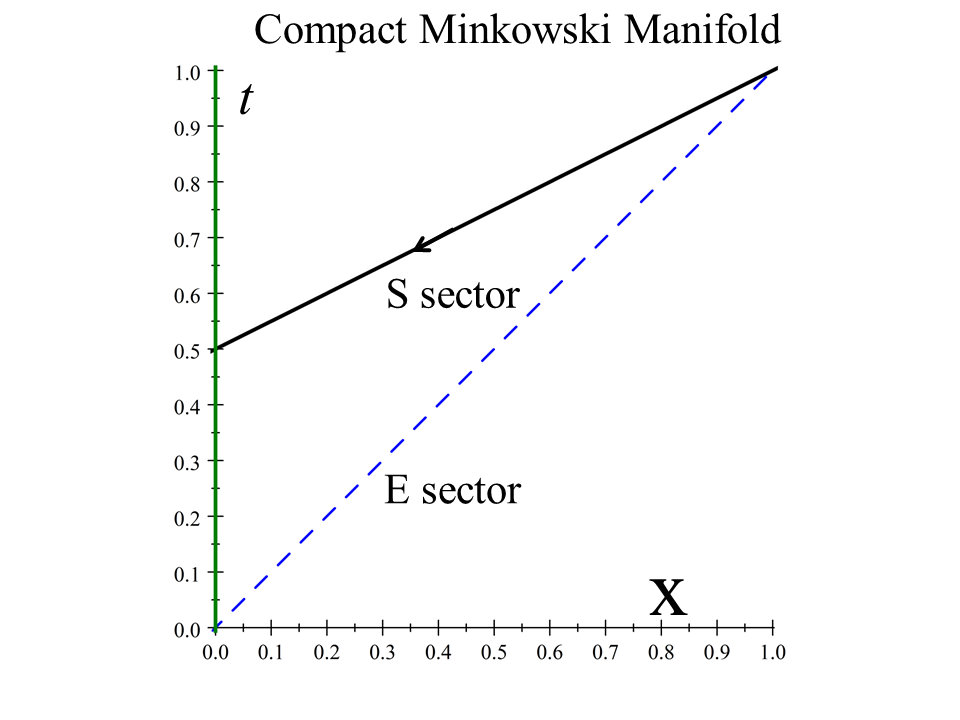} 
\caption{The green line on the \(t\)-axis represents Eq. (\ref{c26b}), while the 
black line corresponds to Eq. (\ref{c26a}). The dashed blue line divides the 
quadrant into sectors \( S \) and \( E \) at a \( 45^{\circ} \) angle. The 
arrow indicates the direction of decreasing the parameter \( x \),  
from 1 to 0, which signifies from a pure state to a stochastic one.}  
\label{fig13}
\end{figure}

In matrix representation the state (\ref{c0}) is 
\be
{\rho_{D7}}\left( x\right) = \frac{1}{4}\left( 
\begin{array}{cccc}
1-x & 0 & 0 & 0 \\ 
0 & 1+x & -2x & 0 \\ 
0 & -2x & 1+x & 0 \\ 
0 & 0 & 0 & 1-x%
\end{array}%
\right) ,  \label{c20c}
\ee
which is a mixed state for any $x<1$, it is totally stochastic for \(x=0\), and is a pure 
state for $x=1$. The eigenstates and eigenvalues are: 
\begin{equation}
\left\vert \psi _{-}\right\rangle \leftrightarrow \lambda _{s}\left(
x\right) =\frac{1}{4}\left( 3x+1\right) ,  \label{c22}
\end{equation}
for the singlet, and 
\begin{equation}
\left\{ \left\vert 11\right\rangle ,~\left\vert 00\right\rangle ,\left\vert
\psi _{+}\right\rangle \right\} \leftrightarrow \lambda _{t}\left( x\right) =%
\frac{1}{4}\left( 1-x\right) ,  \label{c24}
\end{equation}
for the triplet, displaying a triple degeneracy. The new parameters, which are 
also coordinates in the CMM are 
\begin{subequations} \label{c26}
\begin{eqnarray}
\left( t_{-},u_{-},v_{+},w_{-}\right)  &=&\left( \left(1+x \right)/2,\ 0 ,\ -x ,\ 0 \right), 
\label{c26a} \\
\left( t_{+},u_{+},v_{-},w_{+}\right)  &=&\left( \left( 1-x\right)/2,\ 0,\ 0,\ 0\right) ,
\label{c26b} 
\end{eqnarray}
\end{subequations}
and one defines a fictitious ``light cone'' line at an angle $45^{\circ}$ as 
\be
\left( t,\ u,\ v,\ w\right)_{45^{\circ}} =\left( x,\ 0 ,\ x ,\ 0 \right). 
\label{c26c} 
\ee
solely for the purpose of comparison with SR. In the space-time graph illustrated 
in Fig. \ref{fig13}, the blue dashed line at $45^{\circ}$ in Eq. (\ref{c26c}), divides 
the quadrant 
into two sectors. Both, the green and black solid lines are located in the up sector. 
The inclination of the black line relative to the x-axis is steeper than that of the 
dashed blue line. This suggests that if we consider \(t\) as common time, a point along 
the black line will travel faster than a one moving along the blue line. The states 
evolving along the blue line are referred metaphorically as ``luminal'', which implies that the states 
evolving along the black line are ``superluminal''. Therefore, in this scenario, there is 
no barrier to establishing an upper limit for speed, this behavior remits to the emblematic 
assertion by \cite{EPR} authors, ``spooky action at a distance''. The vertical green line, 
on the ordinate indicates that the system can exist in any mixture state, 
depending only on the pseudo time.

\begin{figure}[hbt]
\centering
\includegraphics[width=3.8in,height=2.6in]{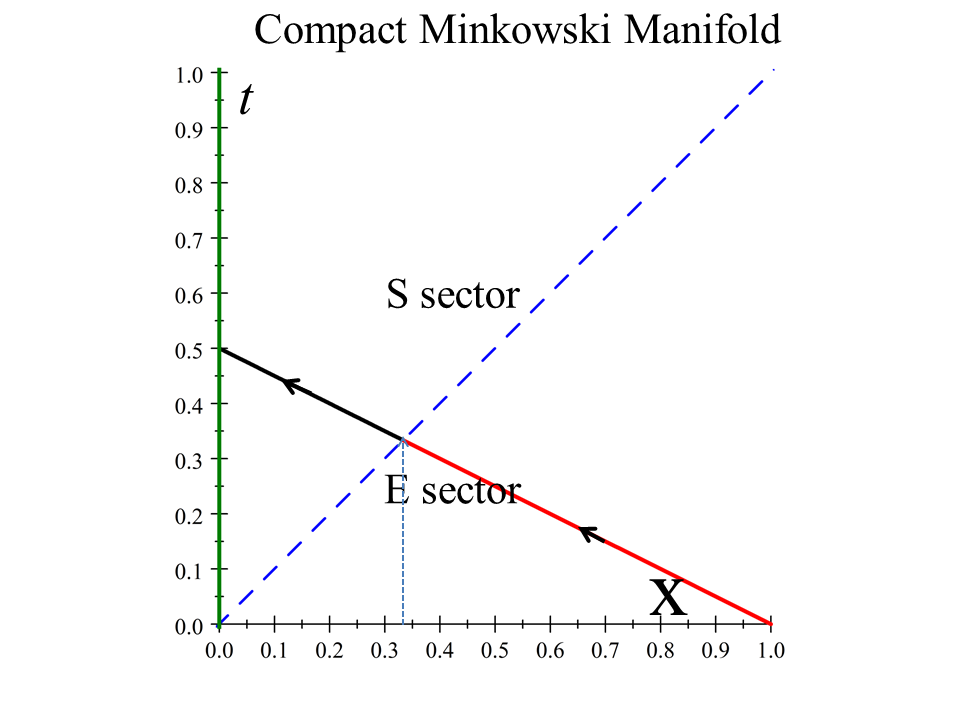} 
\caption{The solid line 
segment in black, in sector $S$ (separable-like), corresponds to Eq. (\ref{c38b}), indicating 
that the qubits are separable for \( x > \frac{1}{3} \). The red line segment in 
sector $E$ (entangled-like) further confirms that the state is entangled. The 
dashed line separates 
separable-like from entangled-like BEW states. The arrows have the same meaning 
as described in the caption of Fig. \ref{fig13}. The green line on the ordinate 
represents Eq. (\ref{c38a}). } 
\label{fig14}
\end{figure}

From the partially transposed matrix
\begin{equation}
\hat{\rho}^{T}_{D7}=\frac{1}{4}\left( 
\begin{array}{cccc}
1-x & 0 & 0 & -2x \\ 
0 & 1+x & 0 & 0 \\ 
0 & 0 & 1+x & 0 \\ 
-2x & 0 & 0 & 1-x%
\end{array}%
\right) ,  \label{c37}
\end{equation}
one gains additional information about the system. Among the eigenstates and eigenvalues, 

\begin{equation}
\left\vert \psi _{+}\right\rangle \leftrightarrow \lambda _{t}\left(
x\right) =\frac{1}{4}\left( 1-3x\right) ,  \label{c23}
\end{equation}
is no more the singlet state, which swaps position with one of the triplets, 
\begin{equation}
\left\{ \left\vert 11\right\rangle ,~\left\vert 00\right\rangle ,\left\vert
\psi _{-}\right\rangle \right\} \leftrightarrow \lambda _{t}\left( x\right) =%
\frac{1}{4}\left( 1+x\right) ,  \label{c25}
\end{equation}
displaying a triple degeneracy. The parameters, which are also coordinates in the 
CMM, are
\begin{subequations}
\label{c38}
\begin{eqnarray}
\left( t_{-},u_{-},v_{-},w_ {+}\right)  &=&\left( \left( 1+x\right) /2,\ 0,\ 0,\ 0\right) ,
\label{c38a} \\
\left( t_{+},u_{+},v_{+},w_{-}\right)  &=&\left( \left( 1-x\right) /2,\ 0,\ -x,\ 0\right) ,
\label{c38b}
\end{eqnarray}
\end{subequations}
Matrix (\ref{c37}) furnishes additional information about the physical system, as the PHC 
can be used here the two sectors displayed in Fig. \ref{fig14} acquire different meaning.
The up sector will characterize the separable BEW states while in the low one will reside 
the entangled states. As such, it is said the dashed line separates separable-like from 
entangled-like sectors for BEW states. In Fig. \ref{fig14}, Eqs. (\ref{c38}) drawn in a 
space-time graph are to be compared with Fig. \ref{fig13}. Here too, for \(t\) 
being the common time, the state evolving along the black line is ``superluminal'', 
still, while residing in the up sector it is separable, and it is surely entangled, 
according to the PHC, when it is found in the down sector.  
%
\subsection{Quadridistance for matrix \texorpdfstring{$\rho_{D7}$}{TEXT}}
\begin{figure}[hbt]
\centering
\includegraphics[width=3.4in,height=2.2in]{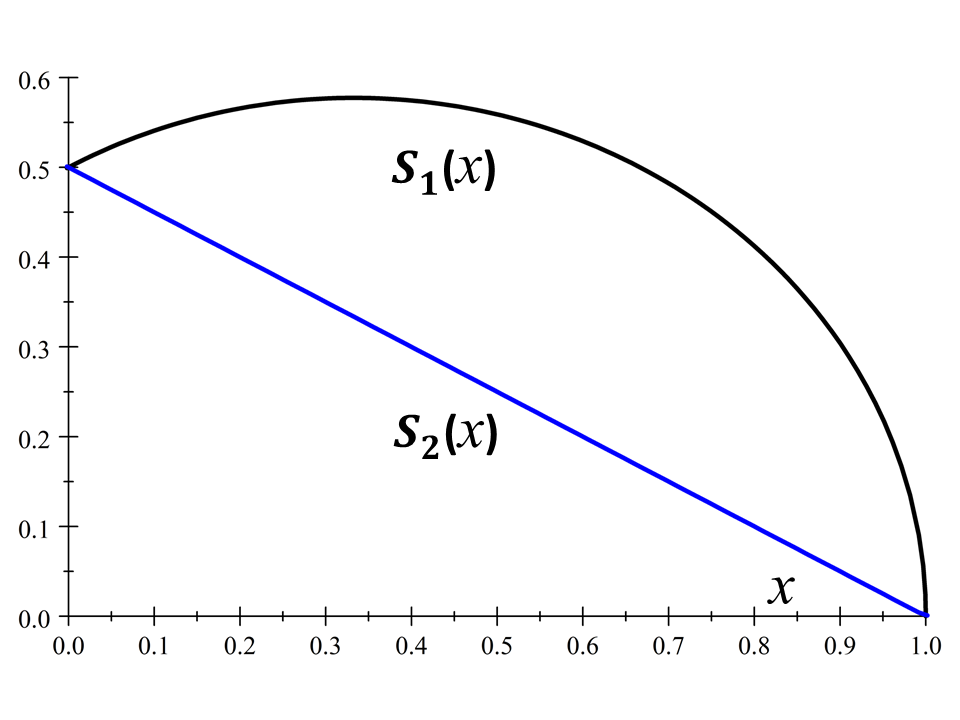} 
\caption{The graphs of \( s_{1}(x) \) and \( s_{2}(x) \) are represented by solid 
lines in black and blue, respectively. When the state is completely stochastic, the 
quadridistances converge to 0.5. Additionally, when the state is pure, 
the values are \( s_{1}(1) = s_{2}(1) = 0 \).}
\label{fig3}
\end{figure}

%
What information do quadridistances 
\begin{subequations} \label{c28}
\begin{eqnarray}
s_{1}^{2}\left( x\right)  &=&t_{-}^{2}-X_{1}^{2} 
=\frac{1}{4}\left(1-x\right) \left( 1+3x\right) \geq 0  \label{c28a} \\
s_{2}^{2}\left( x\right)   &=&t_{+}^{2}-X_{2}^{2} 
=\left( \frac{1-x}{2}\right) ^{2}\geq 0,  \label{c28b} 
\end{eqnarray}
\end{subequations}
provide about the BEW state? In a reference frame in \(\mathcal{R}^3\), the 
distances from the origin are \(X_{1} = 0\) and \(\left|X_{2}\right| = x\). The 
quadridistances follow different paths and have different lengths, with \(s_1(x) > s_2(x)\). 
However, they coincide at two specific points, as illustrated in Fig. \ref{fig3}. 
These coincidences occur when the BEW state is a pure state and when it is stochastic. 

\begin{figure}[hbt]
\centering
\includegraphics[width=3.4in,height=2.2in]{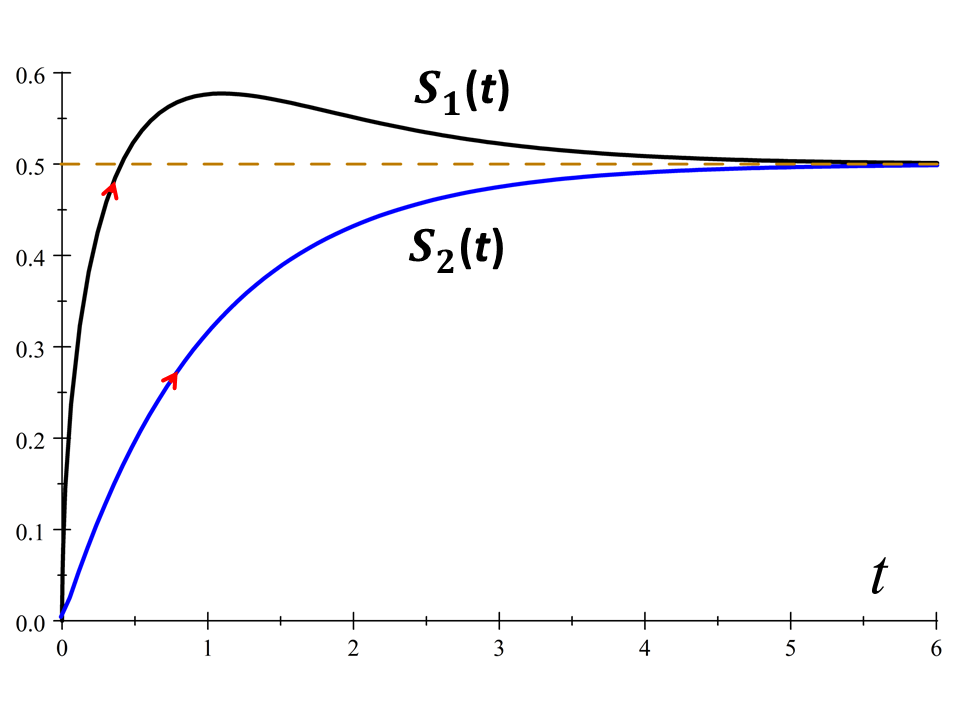} 
\caption{The plots of \( s_{1}(t) \) and \( s_{2}(t) \) are represented by solid lines 
in black and blue, respectively. When the state is fully stochastic, the quadridistances 
converge asymptotically to 0.5. In contrast, when the state is pure, both quadridistances 
equal 0, i.e., \( s_{1}(0) = s_{2}(0) = 0 \). However, despite these endpoints, they 
follow different paths and cover different lengths over the same time intervals.}
\label{fig5}
\end{figure}

Assuming that the 2-qubit state evolves in time (common and also, ``anthropic'' time), intrinsically, 
for example, as $x(t) = \exp(-\gamma t)$, one gets the following expressions for the coordinates:
\begin{subequations} \label{c29}
\begin{eqnarray}
t_{-}&=&\left( 1+\exp \left( -\gamma t\right) \right) /2, \label{c29a} \\
t_{+}&=&\left( 1-\exp \left( -\gamma t\right) \right) /2 \label{c29b} \\
v_{+} &=& -\exp \left( -\gamma t\right) \label{c29c}
\end{eqnarray}
\end{subequations}
Thus the evolution scenario is: at $t = 0$, the qubits are prepared in a singlet state and 
under the effect of the environment, asymptotically ($t\rightarrow \infty$) the state  
acquires a maximally mixed trait, becoming totally stochastic, as expected according to the 
common sense about the ``arrow of time'' concept, the four possible states have the same  
probability to be attained, see Fig. \ref{fig5}, where it was assumed $\gamma =1$.
%
\subsection{Quadridistance for matrix \texorpdfstring{$\rho_{D7}^T$}{TEXT}}
\begin{figure}[hbt]
\centering
\includegraphics[width=3.4in,height=2.2in]{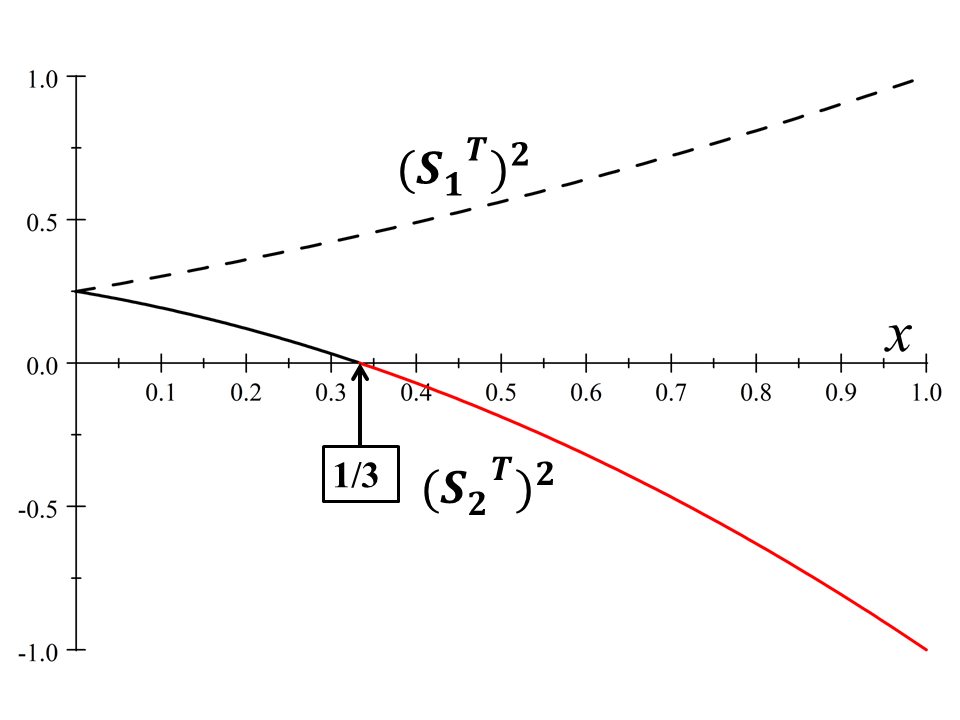} 
\caption{The plot of \((s_{1}^{T})^{2}\) is shown as a dashed line, which does not 
provide any information about entanglement. In contrast, \((s_{2}^{T})^{2}\) is 
represented by a solid line. The black segment that intersects the x-axis at 
\(1/3\) (on the ``light cone'') indicates that the 2-qubit state is separable. 
Meanwhile, the red segment highlights the range of \(x\) values corresponding 
to the entangled 2-qubit state.}
\label{fig01}
\end{figure}

%
The current analysis emphasizes the role of the partial transpose matrix (\ref{c37})
which furnishes complementary and important information about entanglement,
and the position of the entries can be compared with those in matrix (\ref{c20c}). 
The coordinates in the CMM are shown in \ref{c38} and the quadridistances become 
\begin{subequations} \label{c39}
\begin{eqnarray}
s_{1}^{T}\left( x\right) &=&t_{-}=\left( \frac{1+x}{2}%
\right)  =   s_{2} \left( -x\right) ,  \label{c39a} \\
\left( s_{2}^{T}\left( x\right) \right) ^{2} &=&t_{+}^{2}-v_{+}^{2}  
=\frac{1}{4}\left( x+1\right)
\left( 1-3x\right) =  s_{1} ^{2}\left( -x\right) ,   \label{c39b}
\end{eqnarray}
\end{subequations}
where an inner symmetry is revealed when we swap \( x \) with \( -x \). The 
quadridistances derived from the matrix \(\rho_{D7}^T\) change due to the 
reflection resulting from the partial transposition. Consequently, 
\(\left( s_{2}^{T}\left( x\right)\right)^{2}\) takes on negative values 
for \(x \in \left( \frac{1}{3}, 1 \right]\). According to PHC, the qubits are 
entangled when \(x\) is chosen within this interval.  Numerically, 
$s_{1}^{T}\left( 0\right)= s_{2}^{T}\left( 0\right) =1/2$, whereas 
$ s_{1}^{T}\left( 1\right) =1$ and $\left( s_{2}^{T}\left( 1\right)\right)
^{2}=-1$. Thence, 
\begin{equation}
1/2\leq \left.  s_{1}^{T}\left( x\right)\right\vert _{x\in \left[ 0,1%
\right] }\leq 1  \label{c41}
\end{equation}
and
\begin{equation}
-1\leq \left. \left( s_{2}^{T}\left( x\right)\right) ^{2}\right\vert _{x\in \left[ 1,0%
\right] }\leq 1/4.  \label{c42}
\end{equation}
The range of the domain for \(\left( s_{2}^{T}\left( x\right)\right)^{2}\) is broader 
than that for \(\left( s_{1}^{T}\left( x\right)\right)^{2}\), see Fig. \ref{fig01}.
\begin{figure}[hbt]
\centering
\includegraphics[width=3.4in,height=2.2in]{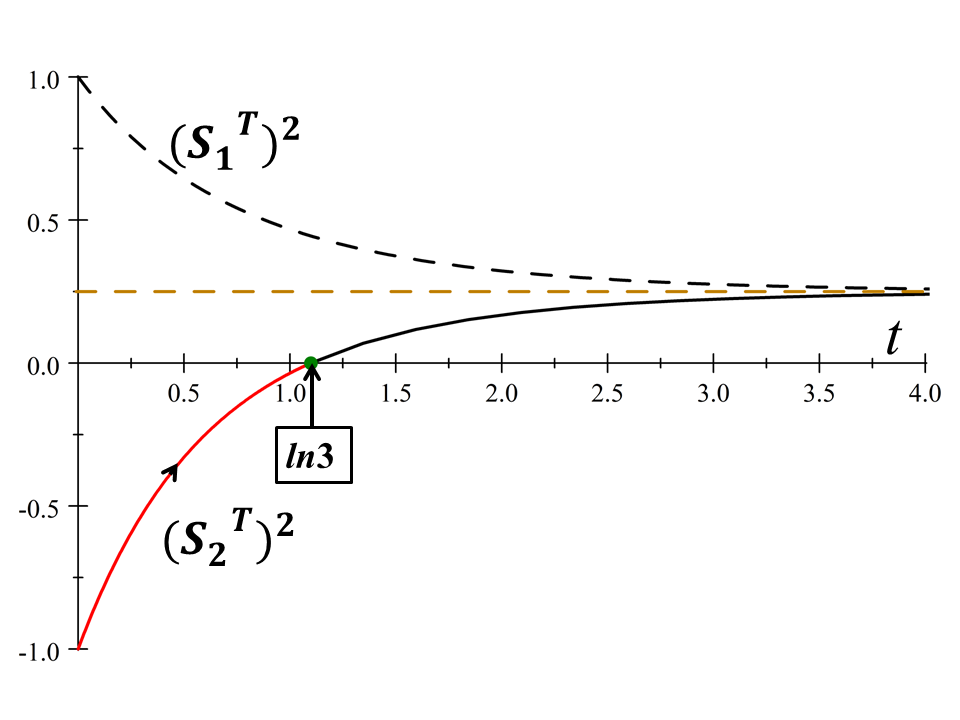} 
\caption{\emph{Evolution in time}: The dashed curve in black represents 
$(s_1^T(x(t)))^2$ as a function of \(t\). In contrast, for $(s_2^T(x(t)))^2$ 
the line segment in red indicates that the qubits are entangled, while the 
black solid line represents the qubits in a separable state. The green dot 
on the $t$-axis at $\ln 3$ is on the ``light cone''. The horizontal dashed 
line in sienna color marks the location of the stochastic state.}
\label{fig2}
\end{figure}

Here too, for a 2-qubit state evolving as \( x(t) = \exp(-\gamma t) \) 
and for the coordinates Eqs. (\ref{c29}), 
as time approaches infinity (\( t \rightarrow \infty \)), 
the state becomes maximally mixed and stochastic. This outcome aligns with 
our intuitive understanding of the concept of the ``arrow of time'', as illustrated 
in Fig. \ref{fig2}.
%
\subsection{Quadrispeed}
%
Concerning the speeds in $\mathcal{R}^3$, from Eqs. (\ref{c26b}) and (\ref{c38a}) one gets
\begin{equation}
V_{1}^{T}=\left|\frac{dv_{-}}{dt_{-}} \right|=\left|\frac{dv_{-}/dx}{dt_{-}/dx}\right|=0  \label{c44}
\end{equation}
and from Eq. (\ref{c38b}) the result is
\begin{equation}
V_{2}^{T}=\left|\frac{dv_{+}}{dt_{+}}\right|=\left|\frac{dv_{+}/dx}{dt_{+}/dx}\right|=2,  \label{c45}
\end{equation}
consequently, the squared quadrispeeds obtain the following values:
\begin{subequations}
\label{c47}
\begin{eqnarray}
\left( \frac{ds_{1}^{T}}{dt_{-}}\right) ^{2} &=&1-\left( V_{1}^{T}\right)
^{2}=1  \label{c47a} \\
\left( \frac{ds_{2}^{T}}{dt_{+}}\right) ^{2} &=&1-\left( V_{2}^{T}\right)
^{2}=-3.  \label{c47b}
\end{eqnarray}
\end{subequations}
The quadrispeed in (\ref{c47a}) does not highlight the dynamics 
of the evolution. In contrast, the quadrispeed (\ref{c47b}) serves as a flag 
indicating that, besides entanglement, the motion is ``superluminal'', euphemistically. Both 
quadrispeeds are independent of \(x(t)\) because the dependence of CMM coordinates 
on \(x\) is linear. However, for other quantum states \cite{gisin,yu-eber1,yu-eber3,almeida},  
that exhibit multi-parametric nonlinear dependence, the quadrispeeds should vary according 
quite differently due to the nature of the parameters. 

Regarding the BEW state, we can associate a different concept 
with ``arrow of time''. Instead of using the time dependence \( x(t) = \exp(-\gamma t) \), 
we could consider \( x(t) = 1 - \exp(-\gamma t) \). This implies that at 
\( t = 0 \), \( x(0) = 0 \), and as \( t \) approaches infinity, 
\( \lim_{t \to \infty} x(t) = 1 \). Consequently, the 2-qubit state that begins 
fully stochastic, under very specific environmental conditions,approaches the pure 
state as time progresses. 

A key distinction between a compact Minkowski manifold derived from a density matrix 
and the manifold of special relativity is that, in real space-time, the coordinates 
\(t, x, y, z\) are assumed to be fundamental in describing the natural world. 
These coordinates do not rely on any underlying parameters, which is not true 
for the compact Minkowski manifold. This differentiation results in the existence 
of a light cone that separates the two regions: time-like and space-like.
%
\section{Summary and conclusion}
%
A longstanding issue in quantum mechanics revolves around the connection between entanglement 
and separability. A system of $N$ qubits, $N\geq2$, can be analyzed in lowest order, as 
$N(N-1)/2$ 2-qubit subsystems, where some or all may hold entanglement.

In the context of the $N(N-1)(N-2)/6$ 3-qubit subsystems, some or all may 
also exhibit entanglement, however, the aim here is to define speed of disentanglement by 
utilizing the appropriate structure of a 2-qubit density matrix for the system of interest. 
A generic $4 \times 4$ matrix contains fifteen free parameters, which can be quite challenging 
to work with. However, by reducing the matrix to a form that depends on only seven free parameters, 
it becomes evident that the matrix can be specified using two polarization vectors (PVs) and a 
$3 \times 3$ correlation matrix (CM). 

According to the symmetries of the state, new parameters can be introduced by reorganizing the 
matrix entries. These parameters indicate that the positivity condition of the state is related 
to quadratic distances in a (3+1)D compact Minkowski manifold.

By applying a local reflection symmetry operation to the 2-qubit matrix, the new matrix 
turns out to be equivalent to one achieved through matrix partial transposition operation. 
This procedure furnishes 
additional information about the system, allowing for the construction of two quadridistances. For 
certain values attributed to the entries, one of the quadridistances may be negative. Hence, according 
to the Peres-Horodecki criterion (PHC), this indicates that the state cannot be expressed in a 
separable form, therefore implying that the qubits are objectively entangled.

In Minkowski's compact manifold, the PHC acquires a geometric interpretation. One can draw trajectories and 
identify two distinct regions: one where qubits are entangled and another where they are 
separable. The boundary between these regions corresponds to the surface of the light cone in special 
relativity. Using similar terminology one region is referred to as ``entangled-like'' 
(analogous to space-like), while the other is ``separable-like'' (similar to time-like).

It maybe not surprising to find a connection between the entanglement of 2-qubit states in Hilbert space and 
the Minkowski manifold of special relativity, even though there may not seem to be a direct link 
between the two. The formalism and examples presented in this essay, along with previous research 
\cite{HSS1,HSS2,MFS}, illustrate some parallelism between these two fields of physics. Notably, 
the speed of disentanglement corresponding to the speed of a superluminal object in special relativity.

According to the current approach, which is supported by the formalism presented here, there is no 
limiting speed for the "spooky action at a distance." 
\appendix
%
\section{Space time geometry: brief reminder} \label{appA}
%
The differential distance between two points in the 3D Euclidean space is expressed as 
a quadratic equation in Cartesian coordinates
\begin{equation}
ds^{2}=dx^{2}+dy^{2}+dz^{2}=dx^{i}g_{ij}dx^{j}  \label{a1}
\end{equation}
where $g_{ij}=\delta _{ij}$ specifies the flat Euclidian metric, which is invariant
under an orthogonal transformation $\mathbb{R}$ of the coordinates, or rotations in 3D,
 $dx^{\prime i}=R^{ik}dx^{k}$ ($\left( R^{T}\right)^{li}R^{ik}=\delta _{kl}$), as $ds^{\prime
2}=R^{ik}dx^{k}g_{ij}R^{jl}dx^{l}=dx^{k}I_{kl}dx^{l}=ds^{2}$, and the unit matrix  
$I_{kl}=\delta _{kl}$. In special relativity (SR) the matrix that specifies the metric 
is pseudo-Euclidean, or hyperbolic, 
\begin{equation}
\mathbf{g}=\left( 
\begin{array}{cccc}
1 & 0 & 0 & 0 \\ 
0 & -1 & 0 & 0 \\ 
0 & 0 & -1 & 0 \\ 
0 & 0 & 0 & -1%
\end{array}%
\right).   \label{a5}
\end{equation}
In the context of differential geometry, the space-time interval between two events is 
expressed as a quadratic equation that includes four terms, representing the quadratic 
infinitesimal geodesic segment
\begin{equation}
ds^{2}=dx_{0}^{2}-d\vec{r}^{\ 2}=dx_{0}^{2}-d\vec{r}\cdot d\vec{r}%
=c^{2}dt^{2}-dx^{2}-dy^{2}-dz^{2}=dx^{\mu }g_{\mu \nu }dx^{\nu }  \label{a7}
\end{equation}
must remain invariant under a space-time transformation: $D^{\alpha \mu} g_{\mu \nu} D^{\nu \beta} 
= g_{\alpha \beta}$, where the indices $\mu, \nu, \alpha, \beta = 0, 1, 2, 3$, with index $0$ 
reserved for time.

For a particle moving along a segment, $ds$, one can define a speed that depends on an internal 
parameter $\vartheta$,
\begin{equation}
u_{g}\left( \vartheta \right) =\frac{ds}{d\vartheta }=+\sqrt{\left( \frac{%
dx_{0}}{d\vartheta }\right) ^{2}-\frac{dx_{i}}{d\vartheta }\frac{dx_{i}}{%
d\vartheta }}  \label{a9}
\end{equation}
If one sets $\vartheta$ as being the common time $t$, then
\begin{equation}
u_{g}\left( t\right) =\sqrt{\left( \frac{cdt}{dt}\right) ^{2}-\frac{dx_{i}}{%
dt}\frac{dx_{i}}{dt}}=\sqrt{c^{2}-\frac{d\vec{r}}{dt}\cdot \frac{d\vec{r}}{dt%
}}=c\sqrt{1-\frac{\vec{v}^{2}}{c^{2}}}.  \label{a13}
\end{equation}
It is assumed that a particle traveling along a geodesic segment has a 
positive speed, $u_{g}(t) > 0$. Therefore, the speed of the particle in three-dimensional 
space should not exceed the speed of light in cosmic vacuum, where $c>v$. However, according 
to equation (\ref{a7}), the inequality $ds^{2} < 0$ is not precluded when a particle is 
moving at a hypothetical superluminal speed with a trajectory drawn  
in the \emph{spacelike} region.
\begin{equation*}
\left( \frac{ds}{dt}\right) ^{2}=\left( \frac{dx_{0}}{dt}\right) ^{2}-\left( 
\frac{d\vec{r}}{dt}\right) ^{2}=c^{2}-\vec{v}^{2}<0 \, .
\end{equation*}
%
\section*{Acknowledgments}
%
Partial support from National Council for Scientific and Technological
Development -- CNPq (Brazil) is acknowledged. 

I thank Dr. Miled Y. H. Moussa, from the Instituto de Física de São Carlos 
for fruitful discussions and suggestions.
%
%

%
\end{document}